\documentclass[twocolumn,showpacs,preprintnumbers,amsmath,amssymb]{revtex4}

\usepackage{graphicx}
\usepackage{dcolumn}
\usepackage{bm}

\usepackage[compact]{titlesec}

\newcommand{\half}{{{\textstyle\frac{1}{2}}}}
\newcommand{\quarter}{{{\textstyle\frac{1}{4}}}}
\newcommand{\be}{\begin{equation}}
\newcommand{\ee}{\end{equation} }
\newcommand{\beqa}{\begin{eqnarray} }
\newcommand{\eeqa}{\end{eqnarray} }
\newcommand{\ba}{\begin{array}}
\newcommand{\ea}{\end{array}}
\newcommand{\bpm}{\begin{pmatrix}}
\newcommand{\epm}{\end{pmatrix}}

\newcommand{\pr}{\prime}

\newcommand{\SU}{\mathbf{SU}}
\newcommand{\Spin}{\mathbf{Spin}}

\newcommand\hcL{{\hat{\cal L}}}

\newcommand{\Off}{\mathbf{O}(4,4)}

\newcommand{\Spinf}{{\Spin(1,3)}}
\newcommand{\oSpinf}{{{\Spin}(3,1)}}

\newcommand\Tr{{\rm Tr}}

\newcommand\cA{{\cal A}}

\newcommand\cD{{\cal D}}

\newcommand\cF{{\cal F}}

\newcommand\cJ{{\cal J}}

\newcommand\cR{{\cal R}}

\newcommand\wcD{\widetilde{\cD}}

\def\tx{\tilde{x}}

\def\na{\nabla}

\def\bra{\bar{a}}
\def\brb{\bar{b}}

\def\bre{\bar{e}}

\def\breta{\bar{\eta}}
\def\bralpha{\bar{\alpha}}
\def\brbeta{\bar{\beta}}
\def\brgamma{\bar{\gamma}}

\def\brpsi{\bar{\psi}}

\def\brl{{\bar{l}}}

\def\brq{{\bar{q}}}

\def\brPhi{{{\bar{\Phi}}}}

\def\brP{\bar{P}}

\def\brV{\bar{V}}

\newcommand\psip{{\psi^{\prime}}{}}
\newcommand{\mbg}{{\mathbf{g}}}

\newcommand{\gYM}{{g_{{\scriptscriptstyle{\cA}}}}}

\begin{document}


\title{Standard Model  as a Double Field Theory}

\author{Kang-Sin Choi}
\email{kangsin@ewha.ac.kr}
\author{Jeong-Hyuck Park}
 \email{park@sogang.ac.kr}

\affiliation{${}^\ast$Scranton Honors Program, Ewha Womans University, Seodaemun-gu, Seoul 120-750, Korea\\
${}^{\dagger}$Department of Physics, Sogang University, Mapo-gu,  Seoul 121-742, Korea}


\begin{abstract}
\noindent We show that, without any extra physical degree introduced,  the Standard Model    can be readily  reformulated as a   Double Field Theory. Consequently, the Standard Model can couple to an arbitrary  stringy  gravitational background in an $\mathbf{O}(4,4)$ T-duality covariant manner and    manifest    two  independent  local Lorentz symmetries,  $\mathbf{Spin}(1,3)\times\mathbf{Spin}(3,1)$. While the diagonal gauge fixing of the twofold spin  groups leads  to the conventional formulation   on the  flat Minkowskian background,  the enhanced symmetry  makes the Standard Model more rigid, and also stringy,  than it appeared. The CP violating  $\theta$-term may no longer be allowed by the symmetry, and hence  the strong CP problem  is  solved.  There are now stronger constraints  imposed   on the possible higher order  corrections.  We speculate  that the quarks and the leptons may possibly belong to the two different spin classes.  
\end{abstract}

                             
                             
\maketitle
Symmetry dictates the structure of the Standard Model~(SM). It  determines the way physical degrees should enter the Lagrangian. The conventional    (continuous)  symmetries of SM   are the Poincar\'{e} symmetry and the $\SU(3)\times \SU(2)\times{\mathbf{U}}(1)$ gauge symmetry. This set of  symmetries  does not forbid the CP violating   $\theta$-term to appear in  the Lagrangian.  But,  there is no experimentally known violation of the CP-symmetry in strong interactions.  This is  the strong CP problem.

Especially when  coupled to gravity, the description of the fermions in SM  calls for  vierbein or tetrad, $e_{\mu}{}^{a}$.  All the     spinorial  indices in the  SM Lagrangian  are then   subject to  the \textit{local} Lorentz  group, $\Spinf$. Physically this gauge symmetry    amounts to  the freedom  to  choose   a locally inertial frame   arbitrarily   at every  spacetime point.  Then to couple the derivatives of  the fermions  to  gamma matrices and to construct a spin connection, it is necessary to employ the tetrad.  
Yet, restricted on the flat Minkowskian background, one can of course choose the  trivial  gauge of the  tetrad, $e_{\mu}{}^{a}{\equiv\delta}_{\mu}{}^{a}$.  However, \textit{a priori}  fermions live on the locally inertial frame and it should be always possible to revive the local Lorentz symmetry and to couple SM to gravity.

Recent development in string theory,  under the name  Double Field Theory~(DFT)~\cite{SiegelDFT1,SiegelDFT2,HZDFT}(\textit{c.f.\,}\cite{review}), generalizes the  Einstein gravity  and reveals   hidden symmetries, such as T-duality.   Especially,   the maximal  supersymmetric   $D{=10}$  DFT has been constructed  to the full order in fermions which exhibits    \textit{twofold}   local Lorentz symmetries~\cite{Jeon:2012hp}(\textit{c.f.~}\cite{Hohm:2011zr}).  Since   the twofold spin   removes the chirality difference between   IIA and IIB,   the  theory  unifies     the IIA and IIB supergravities.  To manifest  all the known as well as the hidden  symmetries simultaneously, it is also necessary to employ  a novel differential geometry which  goes    beyond  Riemann and to some extent    ``Generalized Geometry"~\cite{Hitchin,Gualtieri:2003dx,Coimbra:2011nw}.  

 Physically, the doubling of the local Lorentz symmetries  indicates  a genuine stringy character   that  there are    two separate    locally inertial frames for each   left and   right string mode~\cite{Duff:1986ne}\footnote{The doubling of the local symmetry  group seems to be traced back to Duff~\cite{Duff:1986ne}.  It has been then   rediscovered   in \cite{SiegelDFT2,Jeon:2011cn}, as well as    in the Generalized Geometry  literature~\cite{Coimbra:2011nw}.  }.   
By comparison,  there should be  only one locally inertial frame for a point-particle.

In this Letter,    we  show that,  without introducing any additional  physical degree,  it is possible to reformulate  the Standard Model as a   sort  of  DFT,  such that  it  couples  to the gravitational DFT    and   manifests all the existing  symmetries  at once, which include $\Off$ T-duality, diffeomorphism invariance, and  a pair of local Lorentz symmetries, $\Spin(1,3)\times\Spin(3,1)$.  
    
The enhanced symmetry then seems to forbid the  $\theta$-term, and hence  the strong CP problem can be solved rather   naturally, without introducing  any extra particle, \textit{e.g.~}axions~\cite{Peccei:1977hh,Kim:1979if}. Further, the doubling of the local Lorentz symmetries  has an  immediate phenomenological consequence    that \textit{the spin of the Standard Model can be  twofold}, and every SM fermion  should choose one of the twofold spin groups for its  own spinorial representation.  The enlarged  symmetry also  puts  stronger constraints now  on the possible higher order  corrections.\\

\textit{Convention.}    Unbarred and barred small letters,  $\alpha,\beta,a,b$ and $\bralpha, \brbeta, \bra,\brb$,  are for the  spin groups, $\Spinf$ and $\oSpinf$ respectively: the spinor indices are Greek and  the vector indices are Latin subject to the flat metrics, $\eta_{ab}=\mbox{diag}(-+++)$ and $\breta_{\bra\brb}=\mbox{diag}(+---)$. 

The gamma matrices are accordingly     twofold, 
\be
\ba{ll}
(\gamma^{a})^{\alpha}{}_{\beta}~:&\{\,\gamma^{a}\,,\,\gamma^{b}\,\}=2\eta^{ab}\,,\\
(\brgamma^{\bra})^{\bralpha}{}_{\brbeta}~:&\{\,\brgamma^{\bra}\,,\,\brgamma^{\brb}\,\}=2\breta^{\bra\brb}\,.
\ea
\ee
Capital  Latin letters, $A,B,\cdots$,  denote  the $\Off$ vector indices   which can be freely raised or lowered by the $\Off$ invariant constant metric, $\cJ_{AB}=\tiny{{\bf{\Big(\ba{cc}0&1\\1&0\ea\Big)}}}$. 

 It is crucial to note that different  types of indices cannot be contracted.   This distinction of the indices  will essentially  forbid  many types of  higher order interactions.

\section*{Gravitational DFT: Stringy Differential Geometry}
In order to reformulate SM as a DFT, we   adopt  the stringy differential geometry explored  in  \cite{Jeon:2010rw,Jeon:2011cn}.  
The  key characteristic   is  the \textit{semi-covariant derivative} which can be completely covariantized  by  certain  projections. \\

\textit{Doubled-yet-gauged coordinate system.}   For the  description of the four-dimensional spacetime,  we employ  a     doubled  eight-dimensional  coordinate system, $\{x^{A}\}$,  which  should be yet  gauged   subject  to    an  equivalence relation~\cite{Park:2013mpa,Lee:2013hma}, 
\be
x^{A}~\sim~x^{A}+\varphi_{1}\partial^{A}\varphi_{2}\,.
\label{aCGS}
\ee
Here $\varphi_{1},\varphi_{2}$ denote  arbitrary  fields in DFT. They include all the physical fields, gauge  parameters  and their arbitrary derivatives.   
Each equivalence class or a gauge orbit in the doubled coordinate space  represents a single physical point.  

The equivalence relation~(\ref{aCGS})  is realized     by  enforcing    that,  all the fields in the  DFT     are  invariant under the coordinate gauge symmetry shift~\footnote{For  more conventional  alternative realization of the coordinate gauge symmetry on a string worldsheet, see \cite{Lee:2013hma}.},  
\be
\ba{ll}
\varphi(x+\Delta)=\varphi(x)\,,\quad&\quad\Delta^{A}=\varphi_{1}\partial^{A}\varphi_{2}\,.
\ea
\label{aTensorCGS}
\ee
This invariance  is in fact  equivalent   (\textit{i.e.~}necessary~\cite{Park:2013mpa} and sufficient~\cite{Lee:2013hma})  to the \textit{section condition}~\cite{SiegelDFT2},
\be
\partial_{A}\partial^{A}= 0\quad:\quad
\partial_{A}\partial^{A}\varphi= 0\,,\quad
\partial_{A}\varphi_{1}\partial^{A}\varphi_{2}= 0\,.
\label{aseccon}
\ee
The diffeomorphism on the doubled-yet-gauged spacetime   is generated by a generalized Lie derivative~\cite{HZDFT}, 
\be
\ba{ll}
\hcL_{X}T_{A_{1}\cdots A_{n}}:=&X^{B}\partial_{B}T_{A_{1}\cdots A_{n}}+\omega_{{\scriptscriptstyle{T\,}}}\partial_{B}X^{B}T_{A_{1}\cdots A_{n}}\\
{}&+\sum_{i=1}^{n}2\partial_{[A_{i}}X_{B]}T_{A_{1}\cdots A_{i-1}}{}^{B}{}_{A_{i+1}\cdots  A_{n}}\,,
\ea
\label{hcL}
\ee
where  $\omega_{{\scriptscriptstyle{T\,}}}$ denotes   the weight.  Only the $\Off$ indices are explicitly   denoted  above and other types of  indices are suppressed. 

\textit{Gravitational fields.}  The  whole massless NS-NS sector in  string theory enters  DFT as geometric objects,  in terms of  a  dilaton, $d$, and a pair of vielbeins,   $V_{Aa}$, $\brV_{A\bra}$.   
While  the vielbeins  are   weightless,   the dilaton gives rise to  the $\Off$  invariant integral measure  after exponentiation, $e^{-2d}$,   carrying unit  weight.
The vielbeins  satisfy  four   \textit{defining properties}~\cite{Jeon:2011cn,Jeon:2011vx}:
\be
\ba{ll}
V_{Aa}V^{A}{}_{b}=\eta_{ab}\,,\quad&\quad
\brV_{A\bra}\brV^{A}{}_{\brb}=\breta_{\bra\brb}\,,\\
V_{Aa}\brV^{A}{}_{\brb}=0\,,\quad&\quad V_{Aa}V_{B}{}^{a}+\brV_{A\bra}\brV_{B}{}^{\bra}=\cJ_{AB}\,.
\ea
\label{defV}
\ee
Namely, they  are orthogonal and complete.  The vielbeins  are covariant $\Off$  vectors as their  indices  indicate.    As a solution to  (\ref{defV}),  they can be, if desired,   parametrized     in terms of a pair of  ordinary    tetrads  and a two-form  gauge potential,  in various ways up to $\Off$ rotations and field redefinitions, \textit{e.g.~}(\ref{Vform1}). 
Yet, in the  present covariant formulation, no particular parametrization needs to be assumed. The defining properties~(\ref{defV}) suffice.

 The  vielbeins naturally generate a pair of symmetric, orthogonal and complete two-index projectors,
\[
\ba{ll}
P_{AB}=P_{BA}=V_{A}{}^{a}V_{Ba}\,,\quad&~~
\brP_{AB}=\brP_{BA}=\brV_{A}{}^{\bra}\brV_{B\bra}\,,\\
P_{A}{}^{B}P_{B}{}^{C}=P_{A}{}^{C}\,,\quad&~~
\brP_{A}{}^{B}\brP_{B}{}^{C}=\brP_{A}{}^{C}\,,\\
P_{A}{}^{B}\brP_{B}{}^{C}=0\,,\quad&~~
P_{A}{}^{B}+\brP_{A}{}^{B}=\delta_{A}{}^{B}\,,
\ea
\label{projection}
\]
which, with the dilaton,  $d$, constitute   the ``metric" formulation of the  bosonic DFT (The difference between the projectors is known as ``generalized metric"~\cite{HZDFT}).

\textit{Semi-covariant derivative: complete covariantization.} The semi-covariant derivative~\cite{Jeon:2010rw,Jeon:2011cn} is  defined by 
\be
\ba{ll}
\na_{C}T_{A_{1}A_{2}\cdots A_{n}}:=&\partial_{C}T_{A_{1}A_{2}\cdots A_{n}}-\omega_{{\scriptscriptstyle{T\,}}}\Gamma^{B}{}_{BC}T_{A_{1}A_{2}\cdots A_{n}}\\
{}&+\sum_{i=1}^{n}\,\Gamma_{CA_{i}}{}^{B}T_{A_{1}\cdots A_{i-1}BA_{i+1}\cdots A_{n}}\,.
\ea
\label{asemicov}
\ee
By analogy with Christoffel symbol, the   connection  can be uniquely  chosen~\cite{Jeon:2011cn},
{\small{\be
\ba{ll}
\Gamma_{CAB}=&2\left(P\partial_{C}P\brP\right)_{[AB]}
+2({{\brP}_{[A}{}^{D}{\brP}_{B]}{}^{E}}-{P_{[A}{}^{D}P_{B]}{}^{E}})\partial_{D}P_{EC}\\
{}&-\textstyle{\frac{4}{3}}
(\brP_{C[A}\brP_{B]}{}^{D}+P_{C[A}P_{B]}{}^{D})(\partial_{D}d+(P\partial^{E}P\brP)_{[ED]})\,,
\ea
\label{Gammao}
\ee}}
such that it  satisfies  torsionless conditions,  $\Gamma_{C(AB)}{=0}$, $\Gamma_{[CAB]}{=0}$, and  makes the semi-covariant derivative    compatible with the $\Off$ invariant metric, the projectors and the  dilaton, 
\be
\ba{lll}
\na_{A}\cJ_{BC}=0\,,~&~\na_{A}P_{BC}=0\,,~&~\na_{A}\brP_{BC}=0\,,\\
\multicolumn{3}{c}{
\na_{A}d=-\half e^{2d}\na_{A}(e^{-2d})=\partial_{A}d+\half\Gamma^{B}{}_{BA}=0\,.}
\ea
\label{semicovcomp}
\ee
These are all analogous to the Riemannian Einstein gravity. However, unlike Christoffel symbol, the diffeomorphism~(\ref{hcL}) cannot   transform the connection~(\ref{Gammao}) to vanish point-wise. This can be viewed as the failure of the equivalence principle applied to an extended object, \textit{i.e.~}string.

In order to take care of not only the diffeomorphism~(\ref{hcL}) but also the twofold  local Lorentz symmetries, we need to  further    set   the  \textit{master semi-covariant derivative}~\cite{Jeon:2011vx,Jeon:2011sq},
\be
\cD_{A}:=\na_{A}+\Phi_{A}+\brPhi_{A}
=\partial_{A}+\Gamma_{A}+\Phi_{A}+\brPhi_{A}\,,
\label{mastersemicovD}
\ee
which  includes the spin connections, $\Phi_{A}$,  $\brPhi_{A}$,  for each  local Lorentz group, $\Spinf$, $\oSpinf$ respectively. By definition, in  addition to  the dilaton as (\ref{semicovcomp}),  it is compatible with the vielbeins,
\be
\ba{c}
\cD_{A}V_{Ba}=\partial_{A}V_{Ba}+\Gamma_{AB}{}^{C}V_{Ca}+\Phi_{Aa}{}^{b}V_{Bb}=0\,,\\
\cD_{A}\brV_{B\bra}=\partial_{A}\brV_{B\bra}+\Gamma_{AB}{}^{C}\brV_{C\bra}+\brPhi_{A\bra}{}^{\brb}\brV_{B\brb}=0\,.
\ea
\ee
The spin connections are then fixed by the diffeomorphism connection, $\Gamma$~(\ref{Gammao}),
\be
\ba{ll}
\Phi_{Aab}=V^{B}{}_{a}\na_{A}V_{Bb}\,,\quad&\quad
\brPhi_{A\bra\brb}=\brV^{B}{}_{\bra}\na_{A}\brV_{B\brb}\,.
\ea
\label{PhibrPhi}
\ee
The master semi-covariant derivative is also compatible with all the constant metrics, $\cJ_{AB},\eta_{ab},\breta_{\bra\brb}$, as well as  the gamma matrices,  $(\gamma^{a})^{\alpha}{}_{\beta},(\brgamma^{\bra})^{\bralpha}{}_{\brbeta}$. Consequently, 
the standard  relation between  the spinorial and the vectorial representations of the spin connections holds, such as $\Phi_{A}{}^{\alpha}{}_{\beta}=\quarter\Phi_{Aab}(\gamma^{ab})^{\alpha}{}_{\beta}$ and $\brPhi_{A}{}^{\bralpha}{}_{\brbeta}=\quarter\brPhi_{A\bra\brb}(\brgamma^{\bra\brb})^{\bralpha}{}_{\brbeta}\,$.

The characteristic of the (master) semi-covariant derivative is that,  although it may not be fully diffeomorphic   covariant by itself,  it can be \textit{completely covariantized}   after being appropriately contracted  with the projectors or the  vielbeins.  The \textit{completely covariant derivatives}, relevant to the present work,  are from \cite{Jeon:2011cn}, 
\be
\ba{lll}
{\brP}_{C}{}^{D}P_{A_{1}}{}^{B_{1}}\cdots P_{A_{n}}{}^{B_{n}}
\cD_{D}T_{B_{1}\cdots B_{n}}&\Longleftrightarrow&\cD_{\bra}T_{b_{1}\cdots b_{n}}\,,\\
{P}_{C}{}^{D}\brP_{A_{1}}{}^{B_{1}}\cdots \brP_{A_{n}}{}^{B_{n}}
\cD_{D}T_{B_{1}\cdots B_{n}}&\Longleftrightarrow&\cD_{a}T_{\brb_{1}\cdots \brb_{n}}\,.
\ea
\label{covT}
\ee
Further,  acting on  $\Spinf$ spinor,  $\psi^{\alpha}$  (unprimed), or $\oSpinf$ spinor, $\psip{}^{\bralpha}$ (primed), both of which are  $\Off$ scalars,  the completely covariant Dirac operators,   with respect to both diffeomorphisms and local Lorentz symmetries,  are from \cite{Jeon:2011vx,Jeon:2011sq,Jeon:2012kd}, 
\be
\ba{ll}
\gamma^{a}\cD_{a}\psi=\gamma^{A}\cD_{A}\psi\,,\quad&\quad
\brgamma^{\bra}\cD_{\bra}\psip=\brgamma^{A}\cD_{A}\psip\,.
\ea
\label{covDirac}
\ee
For the full list of the completely covariant derivatives, we refer readers  to \cite{Jeon:2011cn,Jeon:2011vx,Jeon:2012kd}  (\textit{c.f.~}\cite{Hohm:2011si}).\\

\section*{Standard Model  Double Field Theory (SM-DFT)}
The SM-DFT consists of gauge bosons, Higgs, and three generations of  quarks and leptons.  We now turn to their DFT descriptions.

\textit{$\SU(3)\times\SU(2)\times\mathbf{U}(1)$ gauge bosons.}  For each gauge symmetry in SM, we  assign  a Lie algebra valued Yang-Mills potential,    $\cA_{B}$, which should be a diffeomorphism  covariant $\Off$ vector.  

We introduce  \textit{gauged} master semi-covariant derivative, 
\be
\wcD_{B}=\cD_{B}-i\sum_{\cA}\,\cR[\cA_{B}]\,,
\ee
where the sum is over all the gauge symmetries,  and $\cR[\cA_{B}]$ denotes the appropriate representation (depending on quark/lepton, left/right or Higgs)  which also includes  the corresponding  coupling constant, $\gYM$.

We  consider the semi-covariant field strength defined  in terms of the semi-covariant derivative~\cite{Jeon:2011kp}, 
\be
\cF_{AB}:=\na_{A}\cA_{B}-\na_{B}\cA_{A}-i\gYM\left[\cA_{A},\cA_{B}\right]\,.
\ee
Unlike the Riemannian case,  the $\Gamma$-connection inside the semi-covariant derivatives are not canceled out. After, \textit{in fact only after}   contractions  with the `orthogonal' vielbeins,  like (\ref{covT}),   it can be   completely covariantized to take the form~\cite{Jeon:2011kp}:
\be
\cF_{a\brb}=V^{A}{}_{a}\brV^{B}{}_{\brb}\cF_{AB}\,.
\label{comcovF}
\ee
Carrying no $\Off$ index,  $\cF_{a\brb}$ is  a diffeomorphism scalar. The Yang-Mills gauge symmetry is realized by
\be
\ba{ll}
\cA_{A}~\rightarrow~\mbg\cA_{A}\mbg^{-1}-i\frac{1}{\gYM}(\partial_{A}\mbg)\mbg^{-1}\,,&~\cF_{a\brb}~\rightarrow~\mbg\cF_{a\brb}\mbg^{-1}\,.
\ea
\label{YMgauge}
\ee
It is crucial to note  that the completely covariant Yang-Mills field strength, $\cF_{a\brb}$~(\ref{comcovF}),   carries  `opposite' vector indices, one unbarred $\Spinf$   and the other barred $\oSpinf$. Clearly then, with   $\cF_{a\brb}$ alone, it is impossible to write the topological $\theta$-term. Hence,  the strong CP problem  is  naturally solved within the above  DFT setup.    On the other hand, the kinetic term of the gauge bosons, along with that of the Higgs and its potential,   read
\be
\sum_{\cA}\Tr(\cF_{a\brb}\cF^{a\brb})-(P^{AB}-\brP^{AB})(\wcD_{A}\phi)^{\dagger}\wcD_{B}\phi\,-\,V(\phi)\,.
\label{gbH}
\ee
It is worth while to note that $P^{AB}-\brP^{AB}$  above  corresponds to   the so-called  ``generalized metric".

Apparently  there appear doubled off-shell degrees of freedom  in the eight-component gauge potential. In order to halve them,  as  in \cite{Berman:2013cli,Hohm:2013jma,Lee:2013hma}, we  may  impose the following ``gauged"  section   condition: 
\be
(\partial_{A}-i\cA_{A})(\partial^{A}-i\cA^{A})=0\,,
\label{secconA}
\ee
which, with  (\ref{aseccon}), implies $\cA_{A}\partial^{A}{=0}$, $\partial_{A}\cA^{A}{=0}$, $\cA_{A}\cA^{A}{=0}$.
For consistency,  this  condition is  preserved   under all the symmetry transformations: $\Off$ rotations, diffeomorphism~(\ref{hcL}) and the Yang-Mills gauge symmetry~(\ref{YMgauge}), see \footnote{Explicitly we note  $[\mbg\cA_{A}\mbg^{-1}-i\frac{1}{\gYM}(\partial_{A}\mbg)\mbg^{-1}]\partial^{A}=0$ and 
$(\hcL_{X}\cA_{A})\partial^{A}=[X^{B}\partial_{B}\cA_{A}+(\partial_{A}X^{B}-\partial^{B}X_{A})\cA_{B}]\partial^{A}=0\,$. } for  demonstration.

\textit{Quarks and leptons.} Since the spin group is twofold,  we need to decide which spin class  each lepton and quark belongs to. The non-diagonal Yukawa couplings to the Higgs doublet   suggest  that all the quarks should belong to the same spin class. This is  separately  true for   the  leptons as well. Hence,    there are two logical possibilities:  the leptons and the quarks share the same spin group, or they belong to the two distinct spin classes.   Yet,  the absence of the  experimental evidence  of the proton decay might  indicate that  they might  belong to the different spin classes.  In this case, the quarks and the leptons  enter the SM-DFT Lagrangian through the kinetic terms as well as the Yukawa couplings to Higgs as follows,
\be
\sum_{\psi}\brpsi\gamma^{a}\wcD_{a}\psi +\sum_{\psi^{\pr}}\brpsi^{\pr}\brgamma^{\bra}\wcD_{\bra}\psi^{\pr}+y_{d}\brq{\cdot\phi}\, d+y_{u}\brq{\cdot\tilde{\phi}}\, u+y_{e}\brl^{\prime}{\cdot\phi}\,e^{\pr}\,,
\label{quarklepton}
\ee
where, without loss of generality, we have assigned the spin group, $\Spin(1,3)$  to the quarks, $\psi{=(}q, u,d)$,   and the other  $\Spin(3,1)$  to the leptons,  $\psi^{\pr}{=(}l^{\pr},e^{\pr})\,$\footnote{Respectively, $q,u/d,l^{\pr},e^{\pr}$ denote the quark doublets, the up/down-type quark singlets, the lepton doublets and the electron-type singlets.  Flavour indices are suppressed.  Further, $\tilde{\phi}$ is the $\SU(2)$ doublet conjugation of the Higgs. }.   Of course, if the quarks and the leptons should belong to the same spin class, we need to remove the primes.

\textit{Higher order corrections.} Possible higher order corrections have been classified in \textit{e.g.~}\cite{Han:2004az,Grzadkowski:2010es}. However, these  did not take  into account the enhanced symmetry we have been considering.    The classification needs to be further constrained. For example,  the completely covariant field strength, $\cF_{a\brb}$~(\ref{comcovF}), is no longer able to  couple to the skew-symmetric bi-fermionic tensors, $\brpsi\gamma^{ab}\psi$ nor $\brpsi^{\pr}\brgamma^{\bra\brb}\psi^{\pr}$,  to form a dimension-$5$ operator. Further, if the quarks and the leptons indeed  belong to the two different spin classes,  one cannot form a bi-fermion, one from a quark and the other from a lepton. One cannot   also contract     a bi-quark  vector with a bi-lepton vector, \textit{i.e.~}$\brpsi\gamma^{a}\psi$ and  $\brpsi^{\pr}\brgamma^{\bra}\psi^{\pr}$, which would form a  dimension-$6$ operator. Experimental observation of this kind of suppression will confirm (or disprove) our conjecture  that the quarks and the leptons may  belong to the two distinct  spin classes.

\textit{Riemannian reduction.}  With the decompositions of the doubled coordinates, $x^{A}=(\tx_{\mu},x^{\nu})$, $\partial_{A}=(\tilde{\partial}^{\mu},\partial_{\nu})$,  and the  gauge potential, $\cA_{A}=(\tilde{A}^{\mu},A_{\nu})$,   we can solve the section conditions~(\ref{aseccon}),  (\ref{secconA})  explicitly.   Up to $\Off$ rotations,  the most general   solution is given by simply setting $\tilde{\partial}^{\mu}\equiv0$ and $\tilde{A}^{\mu}{\equiv0}$, such that $\partial_{A}\equiv(0,\partial_{\nu})$ and $\cA_{A}\equiv(0,A_{\nu})$.    It  is  then  instructive to parametrize the vielbeins in terms of the Kalb-Ramond two-form potential, $B_{\mu\nu}$, and  a pair of tetrads, $e_{\mu}^{~a},\bre_{\mu}^{~\bra}$, as follows~\cite{Jeon:2011cn}
\be
\ba{ll}
V_{Aa}=\textstyle{\frac{1}{\sqrt{2}}}{{\left(\ba{c} (e^{-1})_{a}{}^{\mu}\\(B+e)_{\nu a}\ea\right)}}\,,
&\brV_{A{\bra}}=\textstyle{\frac{1}{\sqrt{2}}}\left(\ba{c} (\bre^{-1})_{\bra}{}^{\mu}\\(B+\bre)_{\nu{\bra}}\ea\right)\,,
\ea
\label{Vform1}
\ee
where  we  set    $B_{\mu a}{=B}_{\mu\nu}(e^{-1})_{a}{}^{\nu}, B_{\mu \bra}{=B}_{\mu\nu}(\bre^{-1})_{\bra}{}^{\nu}$ and 
the twofold tetrads must give the same Riemannian metric  to solve (\ref{defV}):  
$e_{\mu}{}^{a}e_{\nu}{}^{b}\eta_{ab}=-\bre_{\mu}{}^{\bra}\bre_{\nu}{}^{\brb}\breta_{\bra\brb}\equiv g_{\mu\nu}$.   

The above setting  will then, with the price of breaking  the $\Off$ covariance, reduce the SM-DFT characterized by (\ref{gbH}), (\ref{quarklepton}), to    undoubled, \textit{i.e.~}literally four-dimensional, SM equipped with   two copies of the tetrads  which  take  care of the twofold spin  groups  (see \textit{e.g.}~section~3.2 of \cite{Jeon:2011kp}
and Appendix A.4 of \cite{Jeon:2012kd}).  Finally, on the flat background ($B\equiv0$),   the full  gauge fixing, $e_{\mu}{}^{a}{\equiv\delta}_{\mu}{}^{a}, \bre_{\mu}{}^{\bra}{\equiv\delta}_{\mu}{}^{\bra}$,  will completely reduce the SM-DFT to the conventional formulation of SM. \\
\section*{Concluding Remarks}
String theory has not yet  succeeded in deriving  the precise form  of the Standard Model. Yet,  T-duality and the twofold spin structure are  genuine stringy effects,  which even survive~\cite{Cho:2015lha}  after  the Scherk-Schwarz  dimensional reductions of $D{=10}$ DFT~\cite{Aldazabal:2011nj,Geissbuhler:2011mx}.  We have shown that, without any extra physical degree introduced,  the Standard Model    can be readily  reformulated as a   Double Field Theory, such that it can  couple to an arbitrary  stringy  gravitational background in an $\Off$ T-duality covariant manner and    manifest   two  independent  local Lorentz symmetries,  $\Spin(1,3)\times\Spin(3,1)$.  We have further pointed out  the  possibility   that the quarks and the leptons may belong to the two  different  spin classes. The lacking of the experimental observation  of the proton decay seems to support this conjecture. We urge experimentalists to test this.  

Our  formulation  of the Standard Model as a DFT   is limited  to the  classical level.  Exploration of the quantum aspects   require  further analyses.  For example, one might worry about the chiral anomaly cancelation in our formulation of the Standard Model. Since the same gauge potential, $\cA_{B}$   as the   physical quanta,  is minimally coupled  to the quarks and the leptons through  the contractions,  $\cA_{a}{=\cA}_{B}V^{B}{}_{a}$ and $\cA_{\bra}{=\cA}_{B}\brV^{B}{}_{\bra}$,  we expect that  the anomaly cancelation of the triangular Feynman diagrams   should still work.

It will be     interesting to decide the spin class of the dark matter as well.  If the quarks and the leptons should ever share the same spin group, it seems reasonable to expect that  the dark matter spin may be the  different one.   But here we can only speculate.\\

\textit{Acknowledgments.} This work was  supported by   the National Research Foundation of Korea (NRF)   with the Grants,  2012R1A1A1040695, 2012R1A2A2A02046739, 2013R1A1A1A05005747,  2015K1A3A1A21000302.



\end{document}